\documentclass{elsart}

\renewcommand{\aa}{\AA}

\usepackage{citesort}
\usepackage{graphicx}

\usepackage{amssymb}

\begin{document}

\begin{frontmatter}

\title{Influence of subunit structure on the oligomerization state of light
harvesting complexes: a free energy calculation study}

\date{23 March, 2005}

\author[umc]{L.~Janosi},
\author[uci]{H.~Keer},
\author[umc]{I.~Kosztin},
\ead{KosztinI@missouri.edu}
\author[uci]{T.~Ritz\corauthref{cor}}
\corauth[cor]{Corresponding author. Tel.: ++1-949-824-4345}
\ead{tritz@uci.edu}

\address[umc]{Department of Physics and Astronomy, University of Missouri,
  Columbia, MO 65211, USA}
\address[uci]{Department of Physics and Astronomy, University of California, Irvine,
  CA 92697, USA}

\begin{abstract}
Light harvesting complexes 2 (LH2) from {\em Rhodospirillum (Rs.)
molischianum} and {\em Rhodopseudomonas (Rps.) acidophila} form ring
complexes out of eight or nine identical subunits, respectively.
Here, we investigate computationally what factors govern the
different ring sizes. Starting from the crystal structure
geometries, we embed two subunits of each species into their native
lipid-bilayer/water environment. Using molecular dynamics
simulations with umbrella sampling and steered molecular dynamics,
we probe the free energy profiles along two reaction coordinates,
the angle and the distance between two subunits. We find that two
subunits prefer to arrange at distinctly different angles, depending
on the species, at about 42.5$^\circ$ for {\em Rs.~molischianum} and
at about 38.5$^\circ$ for {\em Rps.~acidophila}, which is likely to
be an important factor contributing to the assembly into different
ring sizes. Our calculations suggest a key role of surface contacts
within the transmembrane domain in constraining these angles,
whereas the strongest interactions stabilizing the subunit dimers
are found in the C-, and to a lesser extent, N-terminal domains. The
presented computational approach provides a promising starting point
to investigate the factors contributing to the assembly of protein
complexes, in particular if combined with modeling of genetic
variants.
\end{abstract}

\end{frontmatter}

\section{Introduction}
 In photosynthesis, assemblies of
pigment-protein complexes absorb sunlight and convert its energy
into a biochemical potential. In recent years, tremendous progress
has been made towards identifying the {\em in vivo} structure of the
photosynthetic apparatus, in particular that of purple bacteria
\cite{HU2002,COGD2004,SCHE2004,BAHA2004}. Calculations and
spectroscopic measurements reveal that a close relationship exists
between the efficiency of light harvesting and the geometrical
arrangements of pigment-protein complexes
\cite{FLEM97,SUND99,RITZ2001,RITZ2002A}.

This raises the question as to how nature governs the assembly of
its photosynthetic apparatus within its membrane. The antenna
light-harvesting complex 2 (LH2) of purple bacteria can be
considered a paradigmatic model system to address this question,
because (i) atomic-resolution crystal structures exist for LH2s with
different organizations and (ii) mutagenesis and reconstitution
assays allow for direct experimental studies of key factors in the
assembly of LH2 complexes.

  \begin{figure}[htbp]
  \centering
  \includegraphics[clip,width=2in]{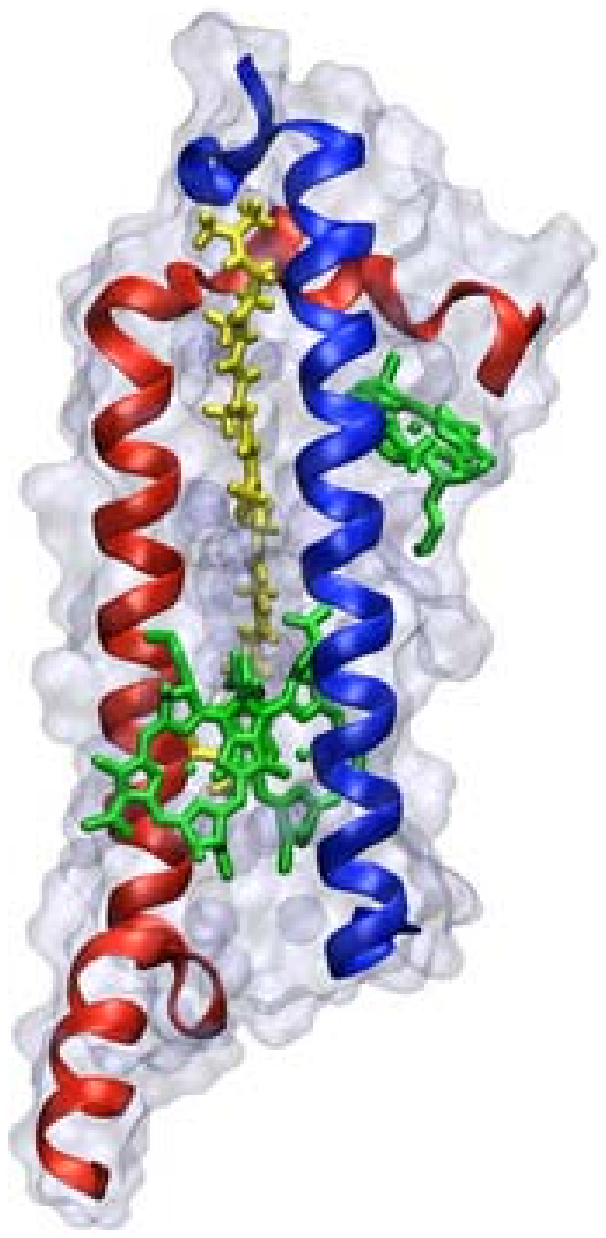}\hspace{4ex}%
  \includegraphics[clip,width=2in]{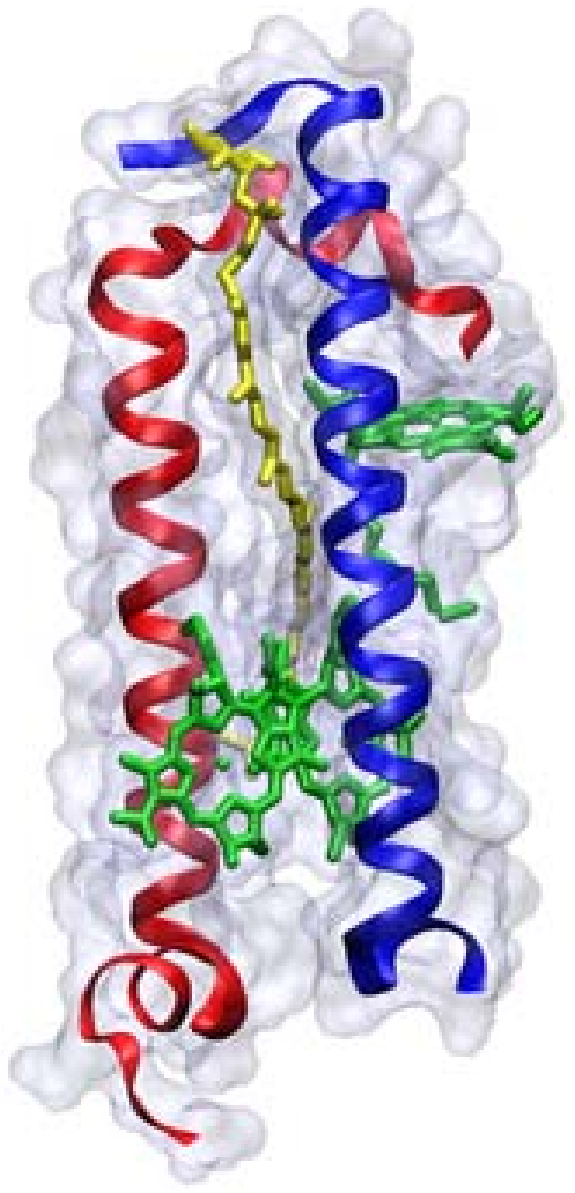}
  \caption{Structure of a subunit of LH2 from {\em Rs.~molischianum}
  (left) and from {\em Rps.~acidophila} (right). Each subunit
  consists of one $\alpha$ polypeptide (blue ribbon), one
  $\beta$-polypeptide (red ribbon), three BChls (green; phytyl chains truncated),
  and one carotenoid (yellow). The N-terminal domains are on top,
  the C-terminal domains on bottom.
  The surface of the subunit is superimposed onto the simplified representations}
  \label{fig:subunits}
\end{figure}

\begin{figure}[htbp]
 \centering
  \includegraphics[clip,height=2.6in]{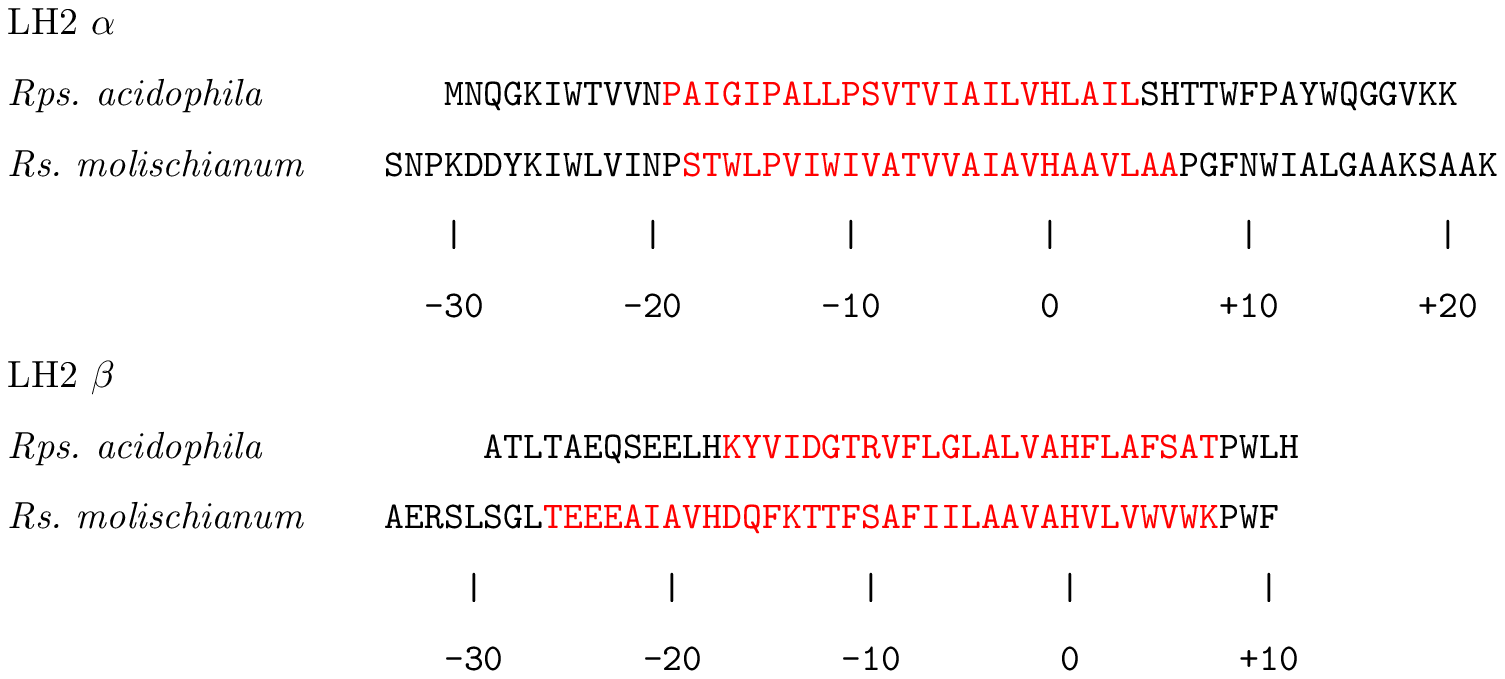}
\caption{Amino acid sequences of the light-harvesting polypeptides
of LH2 from {\em Rs.~molischianum} and {\em Rps.~acidophila}.
Transmembrane helical domains are highlighted in red.}
\label{tab:sequences}
\end{figure}

LH2s are composed of repetitions of a basic unit of two
transmembrane polypeptides, $\alpha$ and $\beta$. Each $\alpha\beta$
heterodimeric subunit non-covalently binds three
bacteriochlorophylls (BChls) and one
carotenoid.Figure~\ref{fig:subunits} shows the basic subunit of LH2
from {\em Rhodospirillum (Rs.) molischianum} and {\em
Rhodopseudomonas (Rps.) acidophila}, respectively, as taken from
their crystal structures \cite{MCDE95,KOEP96}.
Figure~\ref{tab:sequences} shows the corresponding sequences of LH2
$\alpha$ and $\beta$ polypeptides. Both $\alpha$ and $\beta$
polypeptides consist of polar N- and C-termini and a central
hydrophobic region. The N-termini lie on the cytoplasmic side of the
membrane, the C-termini on the periplasmic side. Amino acids in the
central hydrophobic region form two transmembrane $\alpha$-helices.
B850 BChls are ligated to the highly conserved His$_0$ residues near
the C-terminus.

Interestingly, the crystallographic structures of LH2 reveal a
different organization of subunits, a ring of nine
$\alpha\beta$-subunits for {\em Rps.~acidophila}
\cite{MCDE95,PAPI2003}, but of eight $\alpha\beta$-subunits for {\em
Rs.~molischianum} \cite{KOEP96}. EM and AFM studies reveal nonameric
organizations for LH2s from {\em Rhodovulum sufidophilum}
\cite{MONT95}, {\em Rhodobacter (Rb.) sphaeroides}
\cite{WALZ98,SCHE2003}, and {\em Rubrivivax gelatinosus}
\cite{RANC2001,SCHE2001}, whereas a low-light variant form of LH2
from {\em Rps.~palustris} reveals an eight-fold symmetry
\cite{HART2002}. In all of the above cases, the octameric or
nonameric organization appears to be homogeneous within a given
species, although chromatographies for 3D crystallization of LH2
from {\em Rs.~molischianum} suggest the possibility of unstable LH2
forms with different numbers of subunits (H. Michel, personal
comm.). Structural heterogeneity within a species has been described
only for {\em Rs.~photometricum}, where AFM studies suggest that
most LH2s are organized in either eight-, nine-, or ten-fold
symmetry \cite{SCHE2004A}.

Differences in organization have also been reported for the core
light-harvesting complex 1 (LH1), surrounding the photosynthetic
reaction center. LH1, like LH2, is composed of repetitions of an
$\alpha\beta$ heterodimeric subunit; however it contains only two
instead of three BChls per subunit. LH1 complexes were reported to
be formed of 16 $\alpha\beta$-subunits ({\em Rsp.~rubrum}
\cite{JAMI2002}, {\em Blastochloris viridis} \cite{SCHE2003A}), 15
$\alpha\beta$-subunits and one PufX-like peptide {\em
Rps.~palustris} \cite{ROSZ2003}, or 12 $\alpha\beta$-subunits plus
one PufX peptide plus a gap ({\em Rb.~sphaeroides}
\cite{JUNG99,SCHE2004})

Reconstitution assays \cite{PARK88} show that in many cases light
harvesting complexes with very similar optical properties to the
wild-type complexes can be reconstituted {\em in vitro} from their
individual components \cite{TODD98,TODD99}. Truncated versions of
natural proteins, chemically synthesized {\em de novo} proteins, and
mutagenetic gene products have been studied, revealing residues
essential to formation of $\alpha\beta$-subunits and full complexes
\cite{KEHO98,MEAD98,DAVI97,PARK2004,OLSE2003}. A recent study
demonstrated {\em in vivo} assembly of redesigned peptides from {\em
Rb.~sphaeroides} into fully functional light-harvesting complexes
\cite{BRAU2002}

These results suggest that the building blocks of light harvesting
complexes can self-assemble to form stable, functional complexes.
Thus, one should be able to relate the observed differences in
complex organization to the structure of their subunits. What
features of the subunits govern the organization of the complete
ring complexes, in particular the oligomerization state, i.e., the
number of subunits employed in forming a ring?

In the present manuscript, we investigate in how far the variations
in oligomerization states can be explained by changes in the local
interaction angle between neighboring subunits. In theory, one
expects that subunits with a preferred interaction angle of
45$^\circ$ would assemble into a ring of eight subunits ($ 8 \times
45^\circ = 360^\circ$), whereas subunits with a preferred
interaction angle of 40$^\circ$ should form rings of nine subunits.
It appears a remarkable feat for subunits of two helical proteins to
 control a difference in angle as small as 5$^\circ$ in
the presence of conformational fluctuations and solvent dynamics.

Starting from the crystal structures of LH2 from {\em
Rs.~molischianum} and {\em Rps.~acidophila}, henceforth referred to
as MOLI and ACI, respectively, we embed two $\alpha\beta$-subunits
from either species into their native lipid-water environment. We
use molecular dynamics with umbrella sampling and steered molecular
dynamics to probe the free energy surfaces along various reaction
coordinates, namely changes in the angle and distance between the
subunits. The calculation techniques, reviewed in the following
section, represent, arguably, the most accurate calculations
possible on a system of this scale. They provide a reference point
against which results from simpler models can be compared. The
calculations reveal significant differences between subunits from
MOLI and ACI that we discuss with view of the consequences for
protein complex assembly. We discuss future studies necessary to
capture the essential properties governing the oligomerization state
of light harvesting complexes.

\section{Theoretical and Computational Model}
\label{sec:methods}

As a first step in understanding how LH2 subunits assemble into precise ring structures, we
have focused on determining and characterizing the interaction between two LH2 subunits. We
started from a geometry characteristic to a full aggregate. Atomic models for two LH2
subunit dimers, namely for MOLI and ACI, can be readily built starting from their high
resolution crystal structures available from the Protein Data Bank (entry codes 1LGH
\cite{koepke96-581} and 1KZU \cite{prince97-412}, respectively). In order to mimic their
native environment, we have embedded a pair of subunits for both MOLI and ACI into properly
solvated lipid bilayers. The dynamical behavior of the systems were investigated by means of
all atom molecular dynamics (MD) simulations.
Details about the built systems and the MD protocols used are presented at the end of this
section.

Since, presently, MD simulations studies are restricted to the
10-100 nanosecond time scale, they cannot be applied directly to
describe the complete assembly process between two LH2 subunits.
Indeed, lateral diffusion of these protein units in the lipid
membrane occurs on a much longer time scale than the one accessible
by MD simulations. However, suitably designed MD simulations,
combined with statistical mechanical analysis, can be used to
determine the free energy profile or \emph{potential of mean force}
(PMF) \cite{frenkel2002} between the interacting subunits. The PMF
then can be used as input in a suitable stochastic model (e.g.,
Fokker-Planck or Smoluchowski equation \cite{risken96}) for
describing the dynamics of the system at a mesoscopic or even
macroscopic scale.
To this end, one first needs to identify a small number of variables (e.g., distances and/or
angles), hereafter referred to as \emph{reaction coordinates} that describe the relative
separation and orientation of the subunits. Then, one should use a properly designed set of
equilibrium or non-equilibrium MD simulations to calculate the PMF (i.e, free energy) $U(Q)$
of the system as a function of the reaction coordinates $Q$. Once the PMF is determined, the
generalized force exerted between the subunits is equal to $F = -\partial U(Q)/\partial Q$.

In what follows, we define reaction coordinates suitable for
describing the self-assembly of LH2 subunits. Then we briefly
describe two well established methods for calculating PMFs from MD
simulations, namely (i) \emph{umbrella sampling}
\cite{roux95-275,heymann01-1295} combined with the \emph{weighted
histogram analysis method} (WHAM) \cite{kumar92-1011}, and (ii)
\emph{steered molecular dynamics} (SMD)
\cite{isralewitz01-13,grubmuller96-997} in conjunction with the
application of the Jarzynski equality \cite{jarzynski97-2690}.
Method (i) uses equilibrium MD simulations, while (ii) relies on
non-equilibrium simulations \cite{frenkel2002}.

\subsection{Reaction Coordinates $R$ and $\theta$}
\label{sec:RC}

The self-assembly of LH2 rings can be envisioned as a process in
which preformed LH2 subunits are first inserted into the membrane,
then brought within contact distance through diffusion in the lipid
membrane, followed by final locking into the ring specific geometry.
In a first approximation, this process can be modeled as a purely
two dimensional one in which each subunit has a specific anisotropic
2D structure (e.g., a deformed disk) that at the end of the process
forms an N-fold symmetric ring (N=8 for MOLI and N=9 for ACI).

\begin{figure}[htbp]
  \centering
  \includegraphics[width=5in]{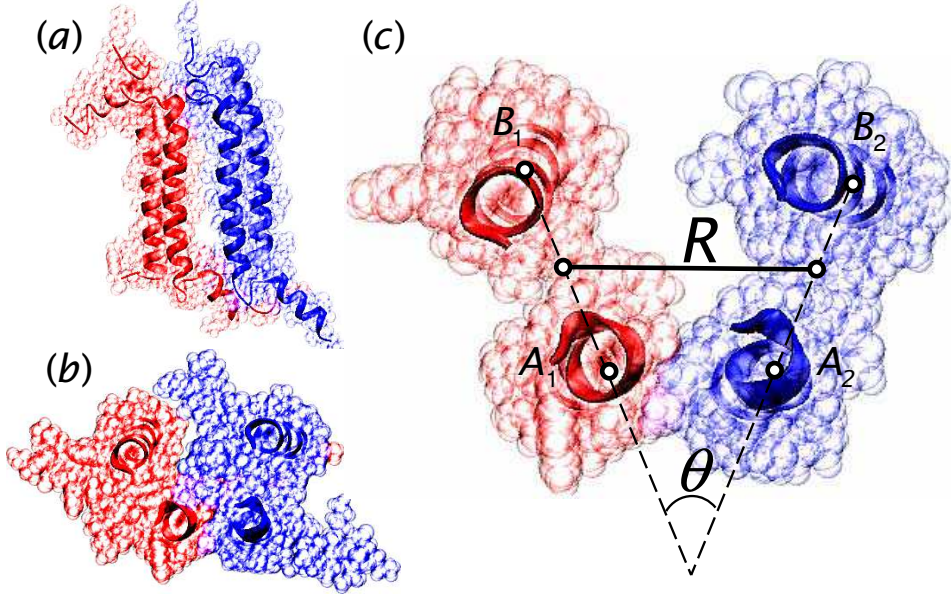}
  \caption{Definition of the reaction coordinates $\theta$ and $R$ illustrated for two subunits of LH2 from {\em Rs.~molischianum}.
  (a) Side view with the N-termini on top.
  (b) Top view from the N-terminal or cytoplasmic side in space-filling representation with transmembrane helices highlighted in
  ribbon representation.
  (c) Top view from from the N-terminal or cytoplasmic side of only the $\alpha$ and $\beta$ polypeptides with definition of reaction coordinates.}
  \label{fig:RC}
\end{figure}

We define two reaction coordinates to characterize the spatial
interaction between LH2 subunits in their native membrane
environment, namely $R$ - the separation between them, and $\theta$
- their relative angular orientation. A more precise definition of
reaction coordinates is illustrated in Fig.~\ref{fig:RC}, displaying
two MOLI subunits (from the equilibrated system) in a side view (a)
and in a top view (b from the N-terminal or cytoplasmic side.
Fig.~\ref{fig:RC}c shows only the transmembrane helical domains of
the two subunits ($A_1$, $B_1$) and ($A_2$, $B_2$), showing a clear
separation of $\alpha$ and $\beta$ polypeptides.

We define the center-of-mass (COM) of the $\alpha$ ($\beta$) apoproteins in subunit $i=1,2$
as $A_i$ ($B_i$). Then $\theta$ by definition is the angle made by the projections of the
vectors $\mathbf{v_i}=\overrightarrow{A_iB_i}$, $i=1,2$, on the $xy$-plane of the membrane.
The separation distance reaction coordinate is defined as the distance between the COMs of
the $\alpha\beta$-apoprotein heterodimer within the $xy$-plane, i.e.,
$R=|\mathbf{R_1}-\mathbf{R_2}|$, where $\mathbf{R_i}$, $i=1,2$ are the projections of the
position vectors of these COMs on the $xy$-plane. Note that, at any instant of time, the
reaction coordinates are determined (through the COMs) by the coordinates and masses of a
selected group of atoms from both subunits.

In selecting the group of atoms that define the reaction coordinates
one has to make sure that under equilibrium (or stationary)
conditions, the reaction coordinates have well defined mean values
and sufficiently small fluctuations. Otherwise, the reaction
coordinates are ill defined and cannot be used to characterize the
system quantitatively. For example, we have found that (see
Sec.~\ref{sec:results}) $R$ and $\theta$ are well defined if one
considers only the heavy atoms in the transmembrane helical (TMH)
domains of the $\alpha\beta$-apoproteins. Adding to the selection
all the heavy atoms in the N- and C-terminal domains, would lead to
substantial increase in the fluctuations of $\theta$ rendering the
latter reaction coordinate meaningless.
Finally, in what follows, we use the convention that $R$ and
$\theta$ refer to particular target values of the reaction
coordinates, while $\tilde{R}\equiv\tilde{R}(\mathbf{q})$ and
$\tilde{\theta}\equiv\tilde{\theta}(\mathbf{q})$ refer to the
instantaneous values of the reaction coordinates determined from the
corresponding positions $\mathbf{q}\equiv\mathbf{q}_j(t)$ of all
defining atoms.

\subsection{Potential of Mean Force $U(\theta,R)$}
\label{sec:pmf}

The PMF $U(Q)\equiv U(\theta,R)$ is the free energy of the system
formed by the two interacting subunits for well defined spatial $R$
and angular $\theta$ separations; for brevity we have introduced the
compact notation $Q\equiv \{\theta,R\}$. In principle, the PMF can
be calculated by integrating out all the degrees of freedom except
for the reaction coordinates \cite{frenkel2002}, i.e.,

\begin{equation}
  \label{eq:pmf1}
  e^{-\beta U(Q)}\equiv p_0(Q)=\int d\Gamma \frac{e^{-\beta H_0(\Gamma)}}{Z_0}
  \delta[Q-\tilde{Q}(\Gamma)] \;,
\end{equation}

where $\beta=1/k_B T$ is the temperature factor ($k_B$ is the
Boltzmann constant and $T$ is the absolute temperature), $p_0(Q)$ is
the equilibrium distribution function of the reaction coordinates,
$H_0(\Gamma)$ is the Hamiltonian of the system as a function of
$\Gamma\equiv \{\mathbf{q,p}\}$ (i.e., all the atomic coordinates
and momenta), $Z_0$ is the partition function and $\delta(x)$ is the
Dirac-delta function whose filtering property guarantees that the
integrand in Eq.~(\ref{eq:pmf1}) is nonzero only when the reaction
coordinates have the requested value, i.e, when
$\tilde{Q}(\Gamma)=Q$. In principle, the equilibrium distribution
function $p_0(Q)$ can be easily computed from MD simulations, since
it is proportional to the binned histogram of the reaction
coordinates calculated along the MD trajectory. Thus, the PMF is
readily given by

\begin{equation}
  \label{eq:pmf0}
  U(Q)=-k_B T\log p_0(Q) \;.
\end{equation}

However, in practice, even the longest equilibrium MD trajectories
will sample only a restricted region of the reaction coordinate
domain (i.e., within the vicinity of the PMF minimum) of interest
and the direct application of Eq.~(\ref{eq:pmf1}) is impractical.

There exist two basic methods for calculating PMFs from MD
simulations widely known as (1) umbrella sampling \cite{roux95-275},
and (2) steered molecular dynamics \cite{park04-5946,hummer01-3658}.
Next we present briefly both methods within a unifying conceptual
framework, and point out under what conditions one should opt for
using one or the other. In both methods a crucial step is to alter
the dynamics of the system by applying a suitable guiding potential.

\subsection{Harmonic Guiding Potential}
\label{sec:hgp}

During equilibrium MD simulations the system explores only a small region of the phase space
$\Gamma$ about the minimum of the sought PMF $U(Q)$. In order to properly sample
energetically more difficult to reach regions, one needs to \emph{guide} or \emph{steer} our
system towards those regions by employing, e.g., a harmonic guiding potential (HGP)
\begin{equation}
  \label{eq:hgp}
  V_Q(\tilde{Q})\equiv V(\tilde{Q}(\Gamma)|Q) = \frac{k_Q}{2}[\tilde{Q}(\Gamma)-Q]^2 \;,
\end{equation}
where $k_Q$ is the stiffness of the HGP. With this extra potential energy, the Hamiltonian
of the new, biased system becomes
\begin{equation}
  \label{eq:hamiltonian}
  H_Q = H_0 + V_Q(\tilde{Q})\;,
\end{equation}
and, as a result, the atoms in the selections that define the
reaction coordinates will experience extra forces

\begin{equation}
  \label{eq:force}
  \mathbf{F}_i = -\frac{\partial{V_Q}}{\partial{\mathbf{r}_i}}=-k_Q
  [\tilde{Q}(\Gamma)-Q] \frac{\partial{\tilde{Q}(\Gamma)}}{\partial{\mathbf{r}_i}}
\end{equation}

Thus, the HGP (\ref{eq:hgp}) will force the system to evolve in the phase space in such a
way that during its time evolution $\tilde{Q}$ will stay confined in the vicinity of
the target $Q$ value.

\subsection{Method I: Umbrella sampling and WHAM}
\label{sec:wham}

In umbrella sampling, one divides the reaction coordinate interval
of interest $(Q_{min},Q_{max})$ that one wants to sample in $N_w$
\emph{windows} by conveniently chosen values $Q_i$,
$i=1,\ldots,N_w$. Next, the reaction coordinate is sampled in each
window separately by preparing identical replicas of the system and
applying the harmonic guiding potential $V_{Q_i}(\tilde{Q})$. As a
result, the biased distribution functions can be readily obtained by
direct sampling of the reaction coordinate for the biased system
\cite{roux95-275,kumar92-1011,frenkel2002}, i.e,
\begin{equation}
  \label{eq:ubs1}
  p_i(Q) = \int d\Gamma \frac{e^{-\beta H_i(\Gamma}}{Z_i}
  \delta[Q-\tilde{Q}(\Gamma)] = \frac{Z_0}{Z_i} e^{-\beta V_i(Q)} p_0(Q)\;,
\end{equation}
where, for brevity, the index $Q_i$ has been replaced by $i$. The
equilibrium distribution in each window is related to the biased
distribution of the reaction coordinate through
\begin{equation}
  \label{eq:ubs2}
  p_0(Q) = \frac{Z_i}{Z_0} e^{\beta V_i(Q)} p_i(Q)\;.
\end{equation}
The standard method to efficiently stitch together the biased $p_i(Q)$'s in order to obtain the
equilibrium $p_0$, and therefore the sought PMF (\ref{eq:pmf0}), is the so called
\emph{weighted histogram analysis method} or WHAM. According to this method, $p_0(Q)$ is
expressed as a weighted sum over the biased distributions $p_i(Q)$ as follows

\begin{equation}
  \label{eq:wham}
  p_0(Q) = \frac{\sum_{i=1}^{N_w} N_i p_i(Q)}{\sum_{i=1}^{N_w} N_i \frac{e^{-\beta V_i(Q)}}{\langle e^{-\beta V_i} \rangle}} \;
\end{equation}
where
\begin{equation}
  \label{eq:wham2}
  \langle e^{-\beta V_i} \rangle = \int dQ p_0(Q) e^{-\beta V_i(Q)} \;.
\end{equation}

The above non-linear coupled WHAM equations, that need to be solved
iteratively, minimize the errors in determining $p_0(Q)$, and
therefore the PMF $U(Q)$. This method will give good PMFs only if
the overlap between windows is substantial. Whenever a reasonable
number of well overlapping windows can be constructed, the umbrella
sampling method with WHAM should be the top choice for calculating
PMFs. In general, this method works very well for determining both
1D and 2D PMFs. We have used this method to calculate the $\theta$
dependent PMF for a constrained separation distance $R$ between LH2
subunits as reported in Sec.~\ref{sec:results}.

\subsection{Method II: Steered Molecular Dynamics and Jarzynski Equality}
\label{sec:smd}

In many cases, it is desirable to vary in time the target value of
the RC according to a prescribed rule. For example, we can guide or
steer our system in the direction of increasing the separation
distance $R$ between the two LH2 subunits during a given simulation
time $\tau$ from an initial value $R_i$ to a final one $R_f$ by
setting in the HGP (\ref{eq:hgp})
\begin{equation}
  \label{eq:smd0}
  Q(t) \equiv R(t) = R_i + v_R t\;, \qquad v_R = (R_f - R_i)/\tau \;.
\end{equation}
By choosing two or more reaction coordinates, one can easily devise
complicated reaction coordinate paths along which the system can be
steered using a moving HGP. With such HGP the Hamiltonian of the
system becomes time dependent, and it can be expressed by inserting
Eq.~(\ref{eq:smd0}) into Eqs.~(\ref{eq:hamiltonian}) and
(\ref{eq:hgp}). Then, the extra forces that need to be applied to
the atoms defining the reaction coordinates can be calculated with
the same formula (\ref{eq:force}). Clearly, the corresponding SMD
simulations are non-equilibrium. Already a few SMD pulling
simulations may help us gain some qualitative insight into the
behavior of the system as it evolves along the reaction coordinate.
We apply this method to investigate the correlation between $\theta$
and $R$ in LH2 subunit dimers when one of the reaction coordinates
is either increased or decreased in time.  The reconstruction of the
PMF $U(Q)$ from such non-equilibrium simulations requires a
sufficiently large number of SMD trajectories. These trajectories
then need to be analyzed by employing the Jarzynski's equality that
connects the free energy differences of two equilibrium states with
the exponential average of the irreversible mechanical work done in
non-equilibrium processes that connect the equilibrium states in
question \cite{jarzynski97-2690,park04-5946,hummer01-3658,gore03-12564}, i.e.,

\begin{equation}
  \label{eq:jarzynski}
  e^{-\beta (F_Q - F_0)} = \langle e^{-\beta W_Q} \rangle\;,
\end{equation}
where the irreversible work along a path that connects states with given RC values $Q_0$ and $Q$
is given by
\begin{equation}
  \label{eq:work}
  W_Q = \int_0^\tau dt \frac{dQ}{dt} \left(\frac{\partial H_Q}{\partial Q} \right)\;.
\end{equation}
Here, $\langle\ldots\rangle$ denotes the average over an ensemble of trajectories.

It should be noted that for very large switching times $\tau$ (adiabatic approximation), the
work becomes reversible and we recover the expected equilibrium result
$W_{\tau\rightarrow\infty}=F_Q-F_0$. On the other hand, for an instantaneous switching time
$W_{\tau\rightarrow 0}=H_Q[\Gamma(0)]-H_0[\Gamma(0)] = \Delta H$, and the Jarzynski equality
(\ref{eq:jarzynski}) yields again the expected result, i.e., $\exp(-\beta \Delta F)=
\langle\exp(\beta \Delta{H}) \rangle$.

In general, the PMF calculation based on the application of the
Jarzynski equality to trajectories from SMD simulations is preferred
to the umbrella sampling method whenever the fluctuations of the
reaction coordinate are small and a huge number of sampling windows
would be required for assuring proper overlap of the reaction
coordinate distribution histograms. However, it is not totally clear
how many SMD trajectories one should use in such a case for
calculating the PMF with sufficient accuracy.
Since for our LH2 subunit dimer system the fluctuations in $R$ are
rather small, i.e., $\sim 0.2~\textrm{\AA}$, and a complete
detachment of the subunits requires an increase in $R$ of
$20-30~\textrm{\AA}$, the present method would be more suitable to
calculate $U(R|\theta)$ than the umbrella sampling method.

\subsection{System modeling and MD simulations}
\label{sec:md}

Here we provide a brief description of the modeling of our MOLI and ACI dimers, and of
the MD simulations performed.

\textit{ACI dimer} -- two adjacent complete LH2 subunits (protein
and cofactors) were extracted from the PDB structure 1KZU
\cite{prince97-412}. After adding the missing hydrogens, the protein
complex was inserted in a pre-equilibrated and hydrated POPC lipid
bilayer. Finally, the system was solvated by adding extra water
layers to the two sides of the lipid bilayer. Besides the proteins
and cofactors the system contained 4014 water molecules and 169 POPC
lipid molecules, with a total system size of 38,594 atoms. The $+4$e
charge of the system was neutralized by properly adding 4 Cl$^{-}$
counter ions. The system was built by using XPLOR
\cite{schwieters03-66} and VMD \cite{humphrey96-33}.

\textit{MOLI dimer} -- using VMD \cite{humphrey96-33} and its
plugins, two complete neighboring units extracted from a fully
equilibrated 8-fold LH2 MOLI ring, reported in one of our previous
MD studies \cite{damjanovic02-031919} were inserted in a
pre-equilibrated and hydrated POPE lipid bilayer, and then solvated
by adding two pre-equilibrated 8~\AA\ thick water layers to each
side of the membrane. The $+4$e charge of the system was neutralized
by properly adding 4 Cl$^{-}$ counter ions. In addition to the
protein and cofactors, the final system contained 8299 water
molecules and 128 POPE lipids, with a total system size of 44,997
atoms.

\textit{Equilibrium MD simulations} -- After proper energy
minimization, the ACI (MOLI) system was equilibrated at 300~K
(310~K), normal atmospheric pressure through a 4~ns (5~ns) long MD
simulation in the NPT ensemble. Periodic boundary conditions and
full electrostatics via the Particle Mash Ewald method were used.
All MD simulations were carried out with the NAMD \cite{kale99-283}
program using the CHARMM~27 parameter set \cite{mackerell98-3586}.
The simulations were carried out on local Linux Beowulf clusters,
and the required time for 1~ns simulation running on 30 CPUs was
about 1.5 days.

\textit{SMD simulations} -- A total of 20 SMD simulations were carried out for each system,
starting from the same state that coincided with the last frame of equilibrium MD run. The
details of the applied harmonic guiding potential for each of the runs are described in
Sec.~\ref{sec:SMD-detach}.

\textit{Umbrella sampling simulations} -- To determine the PMF
$U(\theta|R)$ for $R=R_0=18~\textrm{\AA}$ and
$R=R_x=25~\textrm{\AA}$, umbrella sampling MD runs were carried out
in a a number of windows, starting from $\theta = 33^\circ$ to
$\theta = 53^\circ$, that resulted in well overlapping histograms.
For ACI (MOLI) at $R_0$ the windows were centered around the
following $\theta_i$ angels (measured in degrees): 33, 35, 37, 38,
39, 40, 41, 42, 43, 45, 47, 49, 51 and 53 (35, 36, 37, 38, 39, 40,
41, 42, 43, 45, 47, 49, 51 and 53). For $R_x$ the choice of window
positions were similar. The target angle was enforced by applying a
2D harmonic guiding potential $V_{i,j}(\theta,R)=k_i (\theta -
\theta_i)^2/2 + k_R (R - R_j)^2/2$ with $j=0,x$,
$k_R=80~\textrm{kcal/mol \AA}^2$, and $k_i$ tuned between
$8-10\times 10^3~\textrm{kcal/mol rad}^2$ for achieving optimal
sampling in each window. After exhaustive testings, simulations were
performed for 0.7~ns for each window. Only the last 0.5~ns parts of
the trajectories were used to construct the histograms. Test runs
confirmed that complete sampling in each window is achieved on this
time scale. The $k_i$, $\theta_i$ and $\theta$ histograms have been
used in the WHAM equations to calculate the PMF $U(\theta|R)$ as
described in Sec.~\ref{sec:PMF-theta-R}.

\section{Results}
\label{sec:results}

\subsection{Equilibrium MD simulations of MOLI and ACI dimers}
\label{sec:eq-MD}

\begin{figure}[htbp]
  \centering
  \includegraphics[clip,height=2in]{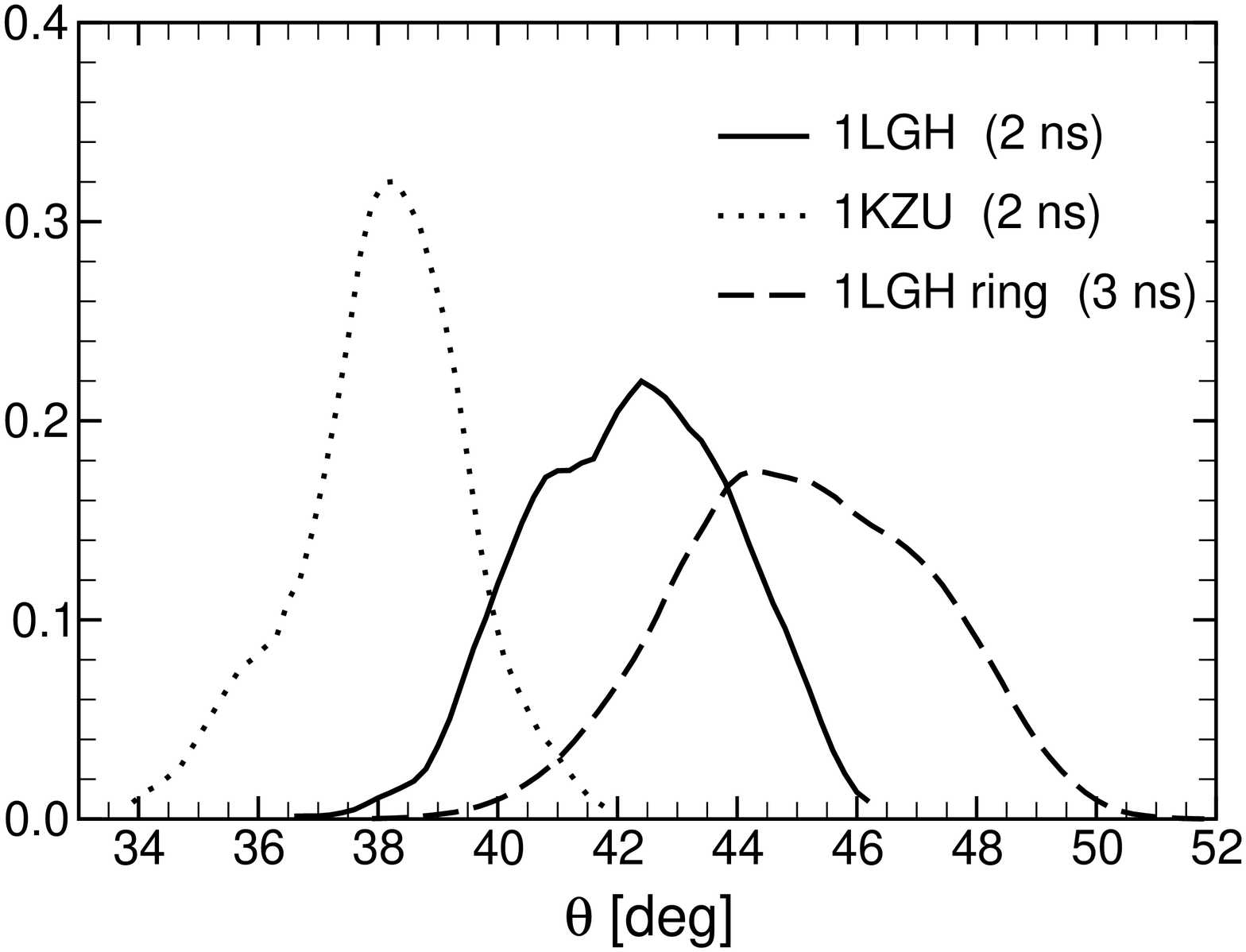}\quad%
  \includegraphics[clip,height=2in]{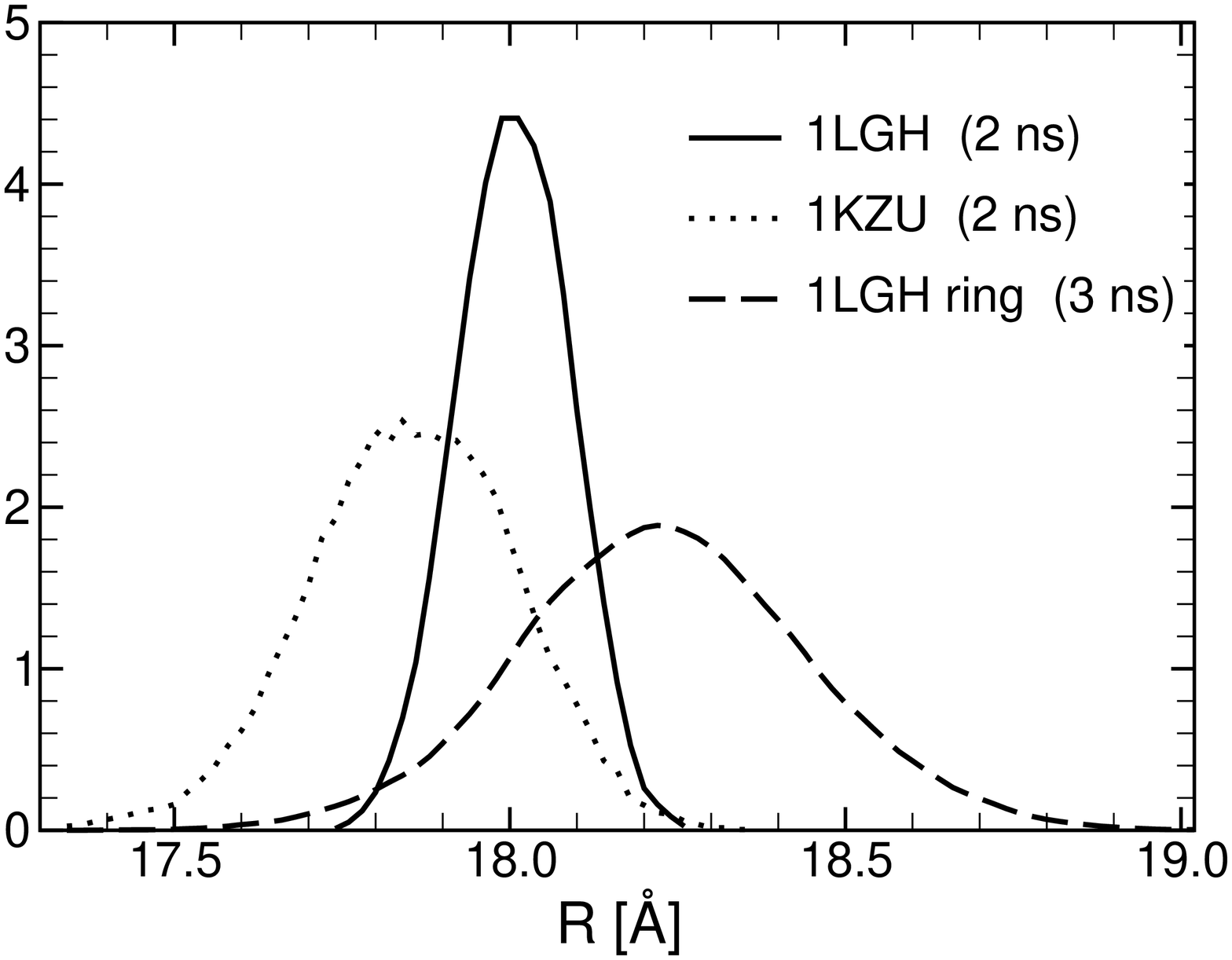}
  \caption{Histograms of the reaction coordinates  $\theta$ (left) and $R$ (right)
    corresponding to equilibrium MD simulations as follows: MOLI dimer, 2~ns run (solid
    curve), ACI dimer, 2~ns run (dashed curve), and MOLI LH2 ring, 3~ns (long-dashed
    curve).}
  \label{fig:H-RC}
\end{figure}

In order to test their reliability and usefulness, we have monitored the time evolution of
the reaction coordinates $R$ and $\theta$, during 5~ns long equilibrium MD simulations of
two LH2 subunit dimers, one from MOLI ({\em Rs.~molischianum}) and one for ACI ({\em
  Rps.~acidophila}).
The simulated systems were prepared as described in Sec.~\ref{sec:methods}.
The calculated values of the reaction coordinates, i.e., $R$ -- the separation, within the
plane of the membrane, between the center-of-mass (COM) of the $\alpha\beta$-polypeptide
pairs in the two subunits, and $\theta$ -- the angle between the projection on the membrane
plane of the lines that connect the COM of the $\alpha$- and $\beta$-polypeptides for each
subunit (cf.~Fig.~\ref{fig:RC}), depends on the selection of atoms used to determine the
COMs. Through extensive testing, we found that the most stable reaction coordinates
correspond to the situation in which only the heavy atoms of the trans-membrane helical
(TMH) domain of the $\alpha\beta$-apoproteins are considered. In this case, the reaction
coordinate assumes well defined mean equilibrium values. The distribution histograms of the
values of the reaction coordinates for the last $2~\textrm{ns}$ of the MD trajectories are
shown in Fig.~\ref{fig:H-RC}.
The peak positions in the histograms (corresponding to the most probable values of the
reaction coordinates) in Fig.~\ref{fig:H-RC} are $R_0 \approx 18\pm 0.08~\textrm{\AA}$,
$\theta_0\approx 42.4^\circ \pm 1.7^\circ$ for MOLI, and $R_0\approx 17.8\pm
0.15~\textrm{\AA}$, $\theta_0\approx 38.1^\circ \pm 1.4^\circ$ for ACI.
For comparison, Fig.~\ref{fig:H-RC} also includes the corresponding
histograms for a complete {\em Rs.~molischianum} LH2 ring, obtained
from the last $3~\textrm{ns}$ part of a $5~\textrm{ns}$ long
equilibrium MD run, and with $R_0\approx 18.2\pm 0.2~\textrm{\AA}$
and $\theta_0= 45\pm 2.1^\circ$. According to these results, we find
that for both MOLI and ACI subunit dimers, $\theta_0$ is about
2-3~$^\circ$ smaller than the compared theoretically expected value
(i.e., 45$^\circ$ for MOLI and 40$^\circ$ for ACI). However, there
is a significant difference in the most probable angle of about
4$^\circ$, close to the theoretically expected difference of
5$^\circ$. For the MOLI LH2 ring $\theta_0$ coincides with the
expected 45$^\circ$, albeit the fluctuations in the angle are
noticeably larger than for the subunit dimers.

Furthermore, fluctuations in $R$ are much smaller than fluctuations
in $\theta$, and the mean value $R_0$ is essentially the same
(within less than half of one \AA) for all three systems.
If one tries to extend the protein atoms selection in the definition
of the reaction coordinates, e.g., by including the CT (C-terminus)
and NT (N-terminus) domains of the $\alpha\beta$-polypeptides, the
resulting reaction coordinates become ill defined especially due to
the sharp increase in the magnitude of the fluctuations. In such
cases, we find that during test MD simulations (not shown) the
fluctuations both in $\theta$ and $R$ increase by more than a factor
of 2.
%
%
These findings are consistent with the knowledge that in general the
TMH regions of membrane proteins are more rigid than the outer
membrane parts.

\subsection{Forced detachment of the subunits using SMD}
\label{sec:SMD-detach}

Next we employ the reaction coordinates introduced above to
investigate the relationship between the spatial separation and
relative orientation of the two LH2 subunits as a result of their
forced detachment. In principle, such a study requires the knowledge
of the 2D potential of mean force (PMF) $U(R,\theta)$, that
describes the statistical mechanical state of the system as a
function of the two reaction coordinates. However, a brute force
determination of the PMF (by direct application of either methods of
calculating PMFs described in Sec.~\ref {sec:methods}) is
computationally prohibitively expensive.

\begin{figure}[htbp]
  \centering
  \includegraphics[clip,width=4.5in]{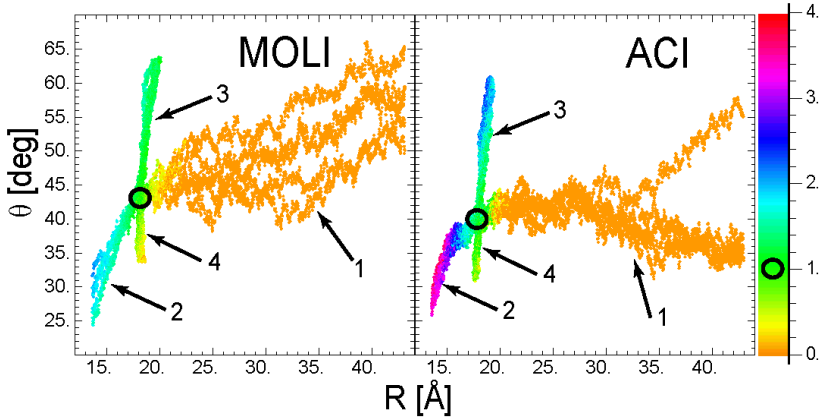}
  \caption{Two dimensional density plot of the volume overlap of the transmembrane protein
    regions of MOLI (left) and ACI (right) dimers along the reaction coordinates $\theta$
    and $R$ recorded in the SMD simulations described in the text; the trajectories
    corresponding to the four different sets of simulations are indicated by numbered
    arrows. The volumes are relative to the value corresponding to the equilibrium reaction
    coordinates $\theta_0$ and $R_0$ (see text), marked by the small circle.}
  \label{fig:overlap}
\end{figure}

Thus, to gain some insight into the mechanism that governs the
relationship between the relative distance and orientation of the
LH2 subunit dimer during the forced separation or compression of the
subunits, we have conducted four sets of 5 SMD simulations each,
starting from initial states that coincide with the last frames of
the equilibrium simulations shown in Fig.~\ref{fig:H-RC}, and
characterized by $R_0=18.2~\textrm{\AA}$ and $\theta_0=43.7^\circ$ for
MOLI and $R_0=18.1~\textrm{\AA}$ and $\theta_0=41^\circ$ for ACI.
These starting values are marked by small circles in
Fig.~\ref{fig:overlap}. In the first (second) set of simulations
$\theta$ was unconstrained while $R$ was increased (decreased) with
a constant rate of $v_R=0.1~$\AA/ps, by applying a harmonic guiding
potential $V(\tilde{R}|R)=k_R (\tilde{R}-R)^2/2$, as described in
Sec.~\ref{sec:methods}, with $k_R=500~\textrm{kcal/mol\AA}^2$.
Similarly, in the third (fourth) set of simulations $R$ was
unconstrained while $\theta$ was increased (decreased) with
$v_{\theta} = 0.1~\textrm{deg/ps}$ through a harmonic guiding
potential $V(\tilde{\theta}|\theta) = k_{\theta} (\tilde{\theta} -
\theta)^2/2$ with $k_{\theta}=5\times 10^4~\textrm{kcal/mol
  rad}^2$. In these potentials, $R$ and $\theta$ are the target values of the reaction
coordinates while $\tilde{R}$ and $\tilde{\theta}$ represent the instantaneous value of the
reaction coordinate as determined from the corresponding atomic coordinates.  The SMD
trajectories in the reaction coordinate plane (i.e., the loci of points with coordinates
\{$R(t), \theta(t)$\}) are shown in Fig.~\ref{fig:overlap} for both MOLI and ACI. The
different trajectories belonging to different sets of simulations are indicated by numbered
arrows.  Trajectory points are color-coded according to the current volume overlap of the TM
parts of the two subunits, normalized to the corresponding initial state value.

The reaction coordinate trajectories exhibit distinctive features
for each set of simulations, with manifest differences between the
two systems.

As soon as the separation $R$ between the two subunits is increased
(set 1) past $R_0$, the TMH domains of the $\alpha\beta$-apoproteins
separate for both MOLI and ACI, an event which is accompanied by a
noticeable increase in the fluctuations of the angle reaction
coordinate.
While in the case of ACI, the separation of the N-terminal and TMH
domains seem to occur simultaneously as $R$ increases, in contrast,
for MOLI, the N-terminal domains of the subunits do not detach until
$R$ becomes larger than 35~\AA. On the other hand, the C-terminal
domains remain connected in both systems even for separations as
large as 40~\AA.
This difference may also be responsible for the profile of the trajectories in set 1. For
MOLI, as $R$ increases, the trajectories cluster in three well distinguished paths, with
occasional partial overlaps and are subject to large $\theta$ fluctuations. This suggests
that the PMF $U(\theta|R)$, for a given $R>R_0$, has a broad global minimum with several (at
least three) local minimums separated by relatively small potential barriers. The location
of the minimum is shifted to $\theta > \theta_0$ values (see below).
On the other hand, for ACI, as $R$ increases, the trajectories remain clustered (albeit with
enhanced $\theta$ fluctuations) suggesting that the PMF $U(\theta|R)$ has a potential well
that is broader than the one for $R_0$.  However, the dramatic deviation of one of the
trajectories from the rest for $R\gtrsim 35~\textrm{\AA}$ suggests that for larger
separation $U(\theta|R)$ may develop a structure with at least two well separated local
minimums, with one equilibrium angle smaller and the other one larger than $\theta_0$.

The behavior of the reaction coordinate trajectories for the
simulations in sets 2, 3 and 4 are qualitatively similar for both
MOLI and ACI. Already a few $\textrm{\AA}$ compression of $R$ (set
2) increases the overlap of the protein subunits several times,
accompanied by a decrease of the angle variable. The trajectories
nicely overlap, indicating that the corresponding PMF $U(\theta|R)$
is similar to the one corresponding to $R_0$, with the minimum
shifted towards a smaller angle than $\theta_0$.

Finally, it is remarkable that for the simulations in sets 3 and 4,
in which $R$ is unconstrained, the latter shows only a slight
increase with respect to $R_0$ as the angle is increased (decreased)
by as much as 20$^\circ$ (10$^\circ$). Thus, based on the results of
our SMD simulations, one may conclude that in general forced
rotation of the relative orientation of the subunits has only very
limited effect on their spatial separation. On the other hand, the
modification of the distance between the subunits in general has
strong impact on the angle between the subunits. This conclusion
also provides support to the notion that the preferred angle between
LH2 subunits is mainly determined by contact interactions between
subunits.
\begin{table}
  \centering
  \begin{tabular}{|l|l|l|l|}
    \hline
    subunit 1&subunit 2&average distance&contact region\\
    residue&residue(s)&for breaking links&\\
    \hline\hline
    $\alpha_1$GLY$_{+8}$  & $\alpha_2$ALA$_{+16}$ &$>$ 40 \AA&C-terminus\\
    $\beta_1$LYS$_{+7}$   &$\alpha_2$ALA$_{+16}$, SER$_{+19}$, ALA$_{+20}$ &$>$ 40 \AA&C-terminus\\
    $\beta_1$PRO$_{+8}$   &$\alpha_2$GLY$_{+15}$, SER$_{+19}$ &$>$ 40 \AA&C-terminus\\
    $\beta_1$TRP$_{+9}$   &$\alpha_2$GLY$_{+15}$ &$>$ 40 \AA&C-terminus\\
    $\alpha_1$PHE$_{+9}$  &$\alpha_2$ILE$_{+12}$ &36 - 38 \AA&C-terminus\\
    $\beta_1$ARG$_{-32}$  &$\beta_2$LEU$_{-27}$  &32 - 35 \AA&N-terminus\\
    $\beta_1$LEU$_{-30}$  &$\alpha_2$SER$_{-18}$ &21 - 22 \AA&N-terminus\\
    $\alpha_1$VAL$_{-22}$ &$\alpha_2$PRO$_{-14}$, SER$_{-18}$ &21 - 24 \AA&TM\\
    \hline
  \end{tabular}
  \caption{Average separation distances at which the most
   important inter-residue links between two LH2 subunits of {\em Rs.~molischianum} are
   broken in SMD simulations.}
  \label{tab:contacts-MOLI}
\end{table}

\begin{table}
  \centering
  \begin{tabular}{|l|l|l|l|}
    \hline
    subunit 1&subunit 2&average distance&contact region\\
    residue&residue(s)&for breaking links&\\
    \hline\hline
    $\alpha_1$THR$_{+7}$  &$\alpha_2$TRP$_{+14}$, GLN$_{+15}$ &$ >$ 40 \AA&C-terminus\\
    $\alpha_1$THR$_{+8}$  &$\alpha_2$GLN$_{+15}$ &$>$ 40 \AA&C-terminus\\
    $\alpha_1$TRP$_{+9}$  &$\alpha_2$TRP$_{+14}$, GLN$_{+15}$ &$>$ 40 \AA&C-terminus\\
    $\beta_1$PRO$_{+8}$   &$\alpha_2$GLN$_{+15}$ &$>$ 40 \AA&C-terminus\\
    $\beta_1$TRP$_{+9}$   &$\alpha_2$TYR$_{+13}$ &34 - 38 \AA&C-terminus\\
    $\alpha_1$VAL$_{-21}$ &$\alpha_2$ALA$_{-18}$, PRO$_{-13}$ &20 - 22 \AA&N-terminus\\
    $\alpha_1$VAL$_{-22}$ &$\alpha_2$ALA$_{-18}$, PRO$_{-19}$ &21 - 27 \AA&N-terminus\\
    $\alpha_1$ALA$_{+2}$  &$\alpha_2$LEU$_{+4}$ &23 - 25 \AA&TM\\
    \hline
  \end{tabular}
  \caption{Average separation distances at which the most
   important inter-residue links between two LH2 subunits of {\em Rps.~acidophila} are
   broken in SMD simulations.}
  \label{tab:contacts-ACI}
\end{table}

Steered molecular dynamics simulations also provide insights into
the key interactions between the subunits: as two subunits are
pulled apart, the links between the subunits break from the weakest
to the strongest. We consider a link broken, if the distance between
the closest contact between two residues becomes larger than 3 \AA.
A list of strongest links, as presented in
Tables~\ref{tab:contacts-MOLI} and \ref{tab:contacts-ACI}, provides
qualitative information on the basis of SMD simulation data. With
increased sampling, more refined analysis in terms of e.g.
interaction energies or free energy barriers would become possible.
%
%
Inspection of the Tables shows that all links in the TMH domain
become separated for distances larger than $R=25~\textrm{\AA}$ for
both cases. In the N-terminal domain, all links become separated for
distances around 25~\AA\ for ACI and around 35~\AA\ for MOLI. This
is consistent with the overlapping volume data shown in
Fig.~\ref{fig:overlap}.
On the other hand, there is one group of links in the C-terminal
domain for both ACI and MOLI that persists even for separations
larger than 40~\AA. This group represents the strongest interactions
conferring overall stability of the binding of the two subunits. An
immediately recognizable difference is that most of the links in ACI
are between the $\alpha$-apoproteins while they are between between
the $\beta 1$ and $\alpha 2$ in MOLI.


\subsection{Calculation of PMF $U(\theta,R)$}
\label{sec:PMF-theta-R}

Our focus in the present manuscript is to investigate the interaction angle between two
subunits. To this end, we have determined two 1D PMFs $U_i(\theta|R_i)$ along $\theta$ by
using \emph{umbrella sampling} and WHAM, as described in Sec.~\ref{sec:methods}.

The calculations were performed for both MOLI and ACI, for two
representative separations, i.e, the equilibrium
$R_0=18~\textrm{\AA}$ and $R_x=25~\textrm{\AA}$. While the choice
for the equilibrium value is obvious, the reason for the $R_x$
selection is that at this particular distance MOLI and ACI are in
qualitatively different separation states. While, for $R_x$, the TMH
domains of the $\alpha\beta$-polypeptides are already separated in
both MOLI and ACI, the N-terminal domain is fully separated only in
ACI, but not in MOLI; the C-terminal domains are still in contact
for both systems.

  \begin{figure}[htbp]
  \centering
  \includegraphics[clip,height=2.3in]{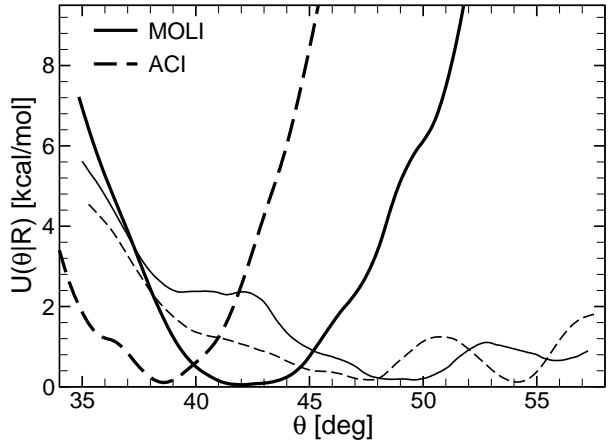}
  \caption{Calculated potentials of mean force $U(\theta|R)$ for
    both MOLI (solid lines) and ACI (dashed lines) dimers.
    The thick (thin) curves correspond to $R=18~\textrm{\AA}$
    ($R=25~\textrm{\AA}$).}
  \label{fig:PMF-R0-Rx}
\end{figure}

The computed PMFs, using umbrella sampling and WHAM, are shown in
Fig.~\ref{fig:PMF-R0-Rx}. For $R_0$, as expected, the PMF for both
MOLI and ACI exhibits a nearly parabolic lineshape with a minimum
that coincides with the position of the peak $\theta_0$ in the
corresponding equilibrium angle distribution histogram in
Fig.~\ref{fig:H-RC}a $\theta_p$. In fact, the PMFs calculated from
those histograms as $U_0(\theta)\propto -k_B T \ln [p_0(\theta)]$
matches rather well the bottom of the full PMF obtained from
umbrella sampling and WHAM (data not shown).
At the equilibrium distance, the PMF of ACI is narrower than that of
MOLI, corresponding to stronger angular constraints for ACI compared
to MOLI. This is consistent with the finding from the SMD
simulations that the fluctuations in the angular reaction coordinate
are smaller for ACI than for MOLI.
Compared to the $R_0$ case, for $R_x$ the PMF for both systems
widens up and acquires additional features. For MOLI, the PMF
exhibits a small plateau at angles around $40^\circ$, and a steep
downhill region for $\theta\gtrsim 42^\circ$ that ends in a broad
minimum around $49^\circ$, followed by a modest potential barrier at
$\sim 53^\circ$.
These features are consistent with the SMD results. Indeed, the angle spread of the SMD
trajectories at $R_x$ extends from $\sim 40^\circ$ to $\sim 55^\circ$. The steep potential
barrier in the PMF at $\lesssim 40^\circ$ explains the lack of trajectory points below this
value. Also, the higher trajectory points density along the plateau region and the broad
minimum is expected.
For ACI, the PMF for $R_x$ shows a $\sim 10^\circ$ wide, rather flat
region starting from $39^\circ$. The fact that the corresponding SMD
trajectory points are clustered only in the interval $38^\circ <
\theta < 44^\circ$ does not conflict with the PMF data but raises
the question why there are no trajectory points up to angles of
$\sim 50^\circ$? There are several possible answers. First, the
small number of SMD trajectories may not provide a proper sampling
of the angles for $R_x$. Secondly, the $\theta$ self diffusion
coefficient may be very small, so that a flat PMF does not lead to
significant dispersion.
The second well in the PMF at $54^\circ$ suggests that eventually a
second branch of trajectories may appear oriented towards higher
angle values as it appears indeed for $R>32~\textrm{\AA}$.

\section{Discussion and Conclusions}

Understanding the assembly of a protein complex requires that one
addresses a set of interrelated questions: (1) What is the temporal
order in which parts are put together? (2) What interactions
stabilize the protein complex or parts of the complex? (3) What
factors govern the (reproducible) stable geometry of the complex?

In this manuscript, we focus on the assembly of a two-subunit
complex from two fully formed $\alpha\beta$-subunits with three BChl
and one carotenoid bound. Spectroscopic observations support the
assumption that this is an important step in the formation of light
harvesting complexes as they show a progression from monomeric BChl
(777 nm) to a BChl dimer bound to an $\alpha\beta$-subunit (820 nm)
to a complex formed of two $\alpha\beta$-subunits (851 nm)
\cite{PAND2001,VEGH2002}. It is, however, unknown whether
carotenoids are already incorporated at this step and how closely
the structure of the $\alpha\beta$-subunit at this point matches the
structure of the $\alpha\beta$-subunit in the crystal structures,
from which we took the coordinates.

An initial analysis of fluctuations in a two-subunit complex
equilibrated in a lipid-water environment revealed that the
hydrophobic core region is rather rigid. Fluctuations in the
centers-of-mass of the core regions in free molecular dynamics runs
are small enough to allow for a meaningful definition of global
coordinates (distance, angle) of the subunits as introduced above.

In order to probe the assembly process, we performed two sets of
calculations with different philosophies. In one set of
calculations, we constrain the subunits at different angles and
allow the system to equilibrate under this constraint. From such
umbrella sampling, we then extract information about the free energy
profile as a function of angle. Pulling two subunits apart through
steered molecular dynamics gives complementary information into key
events of the binding/unbinding process and a rough order of
interaction strengths.

\subsection{Preferred angle}

\begin{figure}[htbp]
    \centering
    \includegraphics[clip,width=6in]{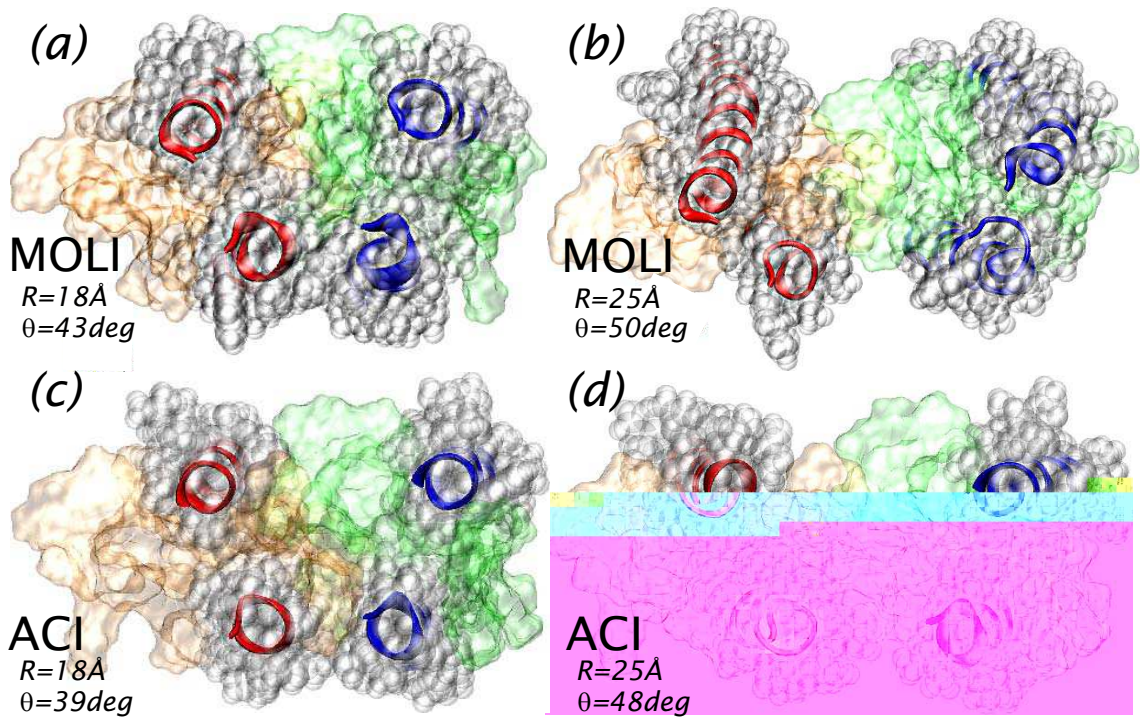}
    \caption{Top view from the N-terminus of the
      transmembrane region of MOLI (a and b) and ACI (c and d) dimers. The values of the
      corresponding reaction coordinate $R$ and $\theta$ are indicated for each case. The
      backbone of the transmembrane helices are shown in cartoon representation. The
      sidegroups of the proteins and the enclosed cofactors (BChls and Car) are shown in
      transparent space-filling representation.}
    \label{fig:R0-Rx}
  \end{figure}

Our calculations show that two subunits in van-der-Waals contact
(center-to-center distance of 18 \AA) arrange at a preferred angle
with each other. Figs.~\ref{fig:R0-Rx}a,c show the transmembrane
domains of subunits in the preferred arrangements at this distance.
For subunits of LH2 from {\em Rs.~molischianum}, the minimum of the
free energy profile (PMF; cf.~Fig.~\ref{fig:PMF-R0-Rx}) is located
at about 42.5$^\circ$, whereas for {\em Rps. acidophila} subunits,
the minimum is located at 38.5$^\circ$. The free energy profiles
closely match parabolic profiles. Whereas changes of about one
degree around the minimum position carry only a small energetic
penalty, it requires about 3 kcal/mol to force two subunits from
{\em Rps.~acidophila} into the angle of 42.5$^\circ$ preferred by
subunits from {\em Rs.~molischianum}. Likewise, about 2 kcal/mol are
required to force two subunits from {\em Rs.~molischianum} into the
angle of 38.5$^\circ$ preferred by subunits from {\em
Rps.~acidophila}. This suggests that the preferred angle between the
subunits plays an important role in guiding the assembly of light
harvesting complexes into a particular ring size or oligomerization
state. However, the preferred angles, while comparing well to the
theoretically expected angles of 45$^\circ$ (8-fold symmetry) and
40$^\circ$ (9-fold symmetry) are somewhat smaller than these. It is
possible that the theoretically expected preferred angle is only
assumed once each subunit is in contact with a subunit on either
side, as would be the case in a ring geometry. The fact that the
average interaction angle in a free molecular dynamics simulation
run of a completed ring is 44.8$^\circ$, much closer to the expected
45$^\circ$, supports this assumption.

To see whether the surface contact between the two subunits is
important in defining a preferred angle, we performed additional
calculations where two subunits are constrained to a
center-to-center distance of 25 \AA. At this distance, subunits of
{\em Rs.~molischianum} are still connected at both their C- and
N-terminal domains, and subunits of {\em Rps. acidophila} are
connected at their C-terminus only. However, the $\alpha\beta$
polypeptides in the transmembrane region are no longer in contact
for either subunit pairs (cf.~Fig.~\ref{fig:R0-Rx}). In this case, the
free energy decreases at larger angles with considerably more
shallow minima than the free energy profiles at the equilibrium
distance of 18 \AA.

These findings suggest that the exact preferred angle between two
subunits is largely defined by the surface interactions in the
transmembrane region. Once the $\alpha\beta$-polypeptides in the
transmembrane region become disconnected, the angular variation
increases significantly (see Fig.~\ref{fig:overlap}). However,
angular constraints, albeit more relaxed, still exist even at larger
separations of the subunits.

Interestingly, the angle between two subunits is more constrained in
{\em Rps. acidophila} than in {\em Rs.~molischianum}. The parabola
of the free energy profile at equilibrium distance (18 \AA) is
narrower for {\em Rps. acidophila} than for {\em Rs.~molischianum}.
Moreover, the five trajectories of {\em Rps. acidophila} during the
steered molecular dynamics simulations show relatively little
angular spread after the transmembrane regions are disconnected and
even after the links in the N-terminal domain are broken. Much more
angular spread can be observed in the case of equivalent steered
molecular dynamics runs for {\em Rs.~molischianum}, even for cases
when the links in both the C- and N-terminal regions are still
intact. As stated above, the fast pulling and small sample size
during this simulation raises a caveat, as the configuration space
may not have been sampled well.

If a more complete sampling of the 2D free energy surface confirms
the observed differences in angular constraints, it would suggest
that in {\em Rps.~acidophila} a motif in the C-terminal domain plays
a role in constraining the angle between subunits in addition to the
hydrophobic surface contacts. One can speculate that this double
constraint of the angle could be the reason why 9-rings are much
more prevalent among LH2 complexes than rings of other sizes.

\subsection{Stabilizing links}

In addition to exploring the factors constraining the angle between
two subunits, the main focus of this publication, we chose
computationally inexpensive SMD calculations to obtain qualitative,
but not quantitative information about the factors stabilizing the
connection between two subunits.

The results from these calculations suggest that links in the
terminal domains play an important role in stabilizing the complex.
These links break last as subunits are pulled apart and thus
represent the strongest interactions between the two subunits. The
strongest interactions are all found in the C-terminal domain.
Interactions in the TMH domain are weakest (cf.~Tables
\ref{tab:contacts-MOLI},\ref{tab:contacts-ACI}).
Only a limited number of experiments address the question of the
requirements for formation of a light harvesting complex; most
mutagenesis experiments are concerned with the formation of a
$\alpha\beta$-subunit. In experimental studies that led to
successful formation of a complete light harvesting complex, the
terminal domains were taken from native sequences, thus offering
little insight into what parts of the terminal domains are required
for successful formation of complete light-harvesting complexes
\cite{TODD98,TODD99,OLSE2003}. Recently, Braun {\em et al.}
demonstrated formation of functional light harvesting complexes from
polypeptides in which all amino acids in the TMH domain except for
the ligating His$_0$ were replaced by alternating pairs of alanin
and leucine residues \cite{BRAU2002}. An addition of four amino
acids in the C-terminal domain resulted in a complete loss of light
harvesting complex formation. This study supports a crucial role the
C-terminal domain in connecting subunits to form complete light
harvesting complexes, whilst indicating a less important role of the
TMH domain.

So far, we have discussed the two dimensions distance and angle
separately from each other. Interestingly, the differences between
{\em Rs.~molischianum} and {\em Rps. acidophila} suggest different
rules by which distances and angles are stabilized. In {\em
Rs.~molischianum}, separation of two subunits appears to require
more energy as the N-terminal domains stay connected far beyond
distances where they are disconnected in {\em Rps. acidophila}.
On the other hand, as discussed above, the angle is more constrained
in {\em Rps. acidophila}. A proper discussion of the interplay
between constraints in angle and distance, requires evaluation of
the full two-dimensional PMF, which is beyond the scope of this
publication.

\subsection{Outlook}
Our results suggest a guided key-lock principle in the assembly of
light harvesting complexes. Interactions in the terminal domains, in
particular the C-terminal domain serve as 'hooks' that connect the
subunits; however, they do not constrain the angle between the
subunits very strongly, although in the case of {\em Rps.
acidophila}, the C-terminus may assist in constraining the angle.
Once the surfaces of the subunits, in particular the BChl
co-factors, start to interlock in the hydrophobic core region, the
angle between the subunits becomes well defined.

To our knowledge, no previous theoretical study has attempted to
address the question as to why some LH2 complexes form 8-rings,
while most form 9-rings. It is therefore a very intriguing result
that we found a distinct difference in the preferred angle between
two subunits that closely matches the expected angle difference of
5$^\circ$ between an 8-ring in {\em Rs.~molischianum} and a 9-ring
in {\em Rps.~acidophila}.

This result is a promising starting point towards understanding the
rules by which light harvesting complexes assemble into rings of
defined sizes. Obviously, many questions remain to be answered, for
example: Do the rules underlying assembly of two subunits remain the
same when many subunits assemble? In other words, are there effects
of different subunit concentrations, leading to phase transitions
between different ring sizes (see e.g., \cite{SCHU2004})? What is
the role of transcription factors and chaperones for assembly {\em
in vivo}?

Genetic manipulation is a powerful tool to obtain information about
the assembly of light harvesting complexes. Several experimental
studies attempting to modify the size of an LH ring by truncating or
swapping C-terminal domains \cite{OLSE2003,RANC2001} did not observe
any alterations of the ring size, although changes in the hydrogen
bonding could be observed. Olsen {\em et al.} speculate that steric
constraints involving BChls, especially B800 BChls play a role in
determining the ring size \cite{OLSE2003}. Our study supports this
view as it suggests an important role of surface contacts in the
hydrophobic region in constraining the angle between two subunits.
Most contacts in the transmembrane region are made by the BChls
ligated to the $\alpha$ and $\beta$ polypeptides. Thus, constraining
the angle between two subunits requires positioning BChls into an
orientation that in turn favors a particular angle between two
subunits due to steric constraints. Amino acid substitutions will
therefore only have an indirect effect on the ring size, by
repositioning the (conserved) BChls into different orientations
through changes in the binding pocket or ligation pattern. It is
therefore by no means straightforward to predict the effects of
amino acid substitutions on the ring size.

Through homology modeling of light-harvesting complexes, one can, in
principle, emulate the genetic manipulation process. Based on such
{\em in silico} mutants and using the calculation techniques
presented here, one can then calculate the preferred interaction
angles of complexes with amino acid deletions or substitutions. The
goal of such calculations would be to predict new complexes for
which, e.g.~a 10-ring or 7-ring architecture is expected. Using
established {\em in vivo} assembly experiments and AFM imaging
techniques, one can test whether these {\em de novo} designs indeed
form the expected ring sizes.

A successful demonstration of the predictive power of molecular
dynamics combined with non-equilibrium techniques could pave the way
towards understanding the principles underlying the assembly of
multimeric membrane protein complexes. Such understanding would have
relevance in controlling nanodevices built from photosynthetic
materials. Furthermore, formation of multimeric protein complexes
occurs in many membrane protein complexes. Combining {\em in silico}
with experimental genetic manipulation could therefore yield
information on the assembly not only of light harvesting complexes,
but of membrane protein complexes in general.

\section{Acknowledgments}

This work was supported by grants from the University of Missouri Research Board (LJ and
IK), the Institute for Theoretical Sciences, a joint institute of Notre Dame University and
Argonne National Laboratory, and the U.S. Department of Energy, Office of Science through
contract No.  W-31-109-ENG-38 (IK and TR).

\bibliographystyle{elsart-num}

\begin{thebibliography}{10}
\expandafter\ifx\csname url\endcsname\relax
  \def\url#1{\texttt{#1}}\fi
\expandafter\ifx\csname urlprefix\endcsname\relax\def\urlprefix{URL }\fi

\bibitem{HU2002}
X.~Hu, T.~Ritz, A.~Damjanovic, F.~Autenrieth, K.~Schulten, Photosynthetic
  apparatus of purple bacteria., Q Rev Biophys 35~(1) (2002) 1--62.

\bibitem{COGD2004}
R.~Cogdell, A.~Gardiner, A.~Roszak, C.~J. Law, J.~Southall, N.~W. Isaacs,
  Rings, ellipses and horseshoes: How purple bacteria harvest solar energy,
  Photosynthesis Research 81 (2004) 207--214.

\bibitem{SCHE2004}
S.~Scheuring, J.~Sturgis, V.~Prima, A.~Bernadac, D.~L\'{e}vy, J.~Rigaud,
  Watching the photosynthetic apparatus in native membranes., Proc. Natl. Acad.
  Sci. USA 101 (2004) 11293--11297.

\bibitem{BAHA2004}
S.~Bahatyrova, R.~Frese, C.~Siebert, J.~Olsen, K.~{van der Werf}, R.~{van
  Grondelle}, R.~Niederman, P.~Bullough, C.~Otto, C.~Hunter, The native
  architecture of a photosynthetic membrane., Nature 430 (2004) 1058--1062.

\bibitem{FLEM97}
G.~Fleming, R.~{van Grondelle}, Femtosecond spectroscopy of photosynthetic
  light-harvesting systems, Curr. Opin. Struct. Biol. 7 (1997) 738--748.

\bibitem{SUND99}
T.~Sundstrom, V.and~Pullerits, R.~{van Grondelle}, Photosynthetic
  light-harvesting: reconciling dynamics and structure of purple bacterial
  {LH2} reveals function of photosynthetic unit, J. Phys. Chem. B 103 (1999)
  2327--2346.

\bibitem{RITZ2001}
T.~Ritz, S.~Park, S.~K., Kinetics of excitation migration and trapping in the
  photosynthetic unit of purple bacteria., Journal of Physical Chemistry B 105
  (2001) 8259--8267.

\bibitem{RITZ2002A}
T.~Ritz, A.~Damjanovic, K.~Schulten, The quantum physics of photosynthesis.,
  Chemphyschem 3~(3) (2002) 243--8.

\bibitem{MCDE95}
G.~{McDermott}, S.~Prince, A.~Freer, A.~{Hawthornthwaite-Lawless}, M.~Papiz,
  R.~Cogdell, N.~Isaacs, Crystal structure of an integral membrane
  light-harvesting complex from photosynthetic bacteria., Nature 374 (1995)
  517--521.

\bibitem{KOEP96}
J.~Koepke, X.~Hu, C.~Muenke, K.~Schulten, H.~Michel, The crystal structure of
  the light-harvesting complex {II} ({B800-850}) from \emph{Rhodospirillum
  molischianum}., Structure 4 (1996) 581--597.

\bibitem{PAPI2003}
M.~Papiz, S.~Prince, T.~Howard, R.~Cogdell, N.~Isaacs, The structure and
  thermal motion of the {B800-850} lh2 complex from rps. acidophila at 2.0 \aa
  resolution and 100 k: new structural features and functionally relevant
  motions., J. Mol. Biol. 326 (2003) 1523--1538.

\bibitem{MONT95}
G.~Montoya, M.~Cyrklaff, I.~Sinning, Two-dimensionalcrystallization and
  preliminary structure analysis of light harvesting {II (B800-850)} complex
  from the purple bacterium \emph{Rhodovulum sulfidophilum}., J. Mol. Biol. 250
  (1995) 1--10.

\bibitem{WALZ98}
T.~Walz, S.~Jamieson, C.~Bowers, B.~P.A., C.~Hunter, Projection structures of
  three photosynthetic complexes from {\em rhodobacter sphaeroides}: {LH2} at 6
  $\aa$, {LH1} and {RC-LH1} at 25 $\aa$, J.Mol.Biol. 282 (1998) 833--845.

\bibitem{SCHE2003}
S.~Scheuring, J.~Seguin, S.~Marco, D.~Levy, B.~Robert, J.~Rigaud,
  Nanodissection and high-resolution imaging of the \emph{Rhodopseudomonas
  viridis} photosynthetic core complex in native membranes by {AFM}., Proc.
  Natl. Acad. Sci. USA. 100 (2003) 1690--1693.

\bibitem{RANC2001}
J.~Ranck, T.~Ruiz, G.~{Pehau-Arnaudet}, B.~Arnoux, F.~{Reiss-Husson},
  Two-dimensional structure of the native light-harvesting complex lh2 from
  \emph{ Rubrivivax gelatinosus} and of a truncated form., Biochim. Biophys.
  Acta. 1506 (2001) 67--78.

\bibitem{SCHE2001}
S.~Scheuring, F.~{Reiss-Husson}, A.~Engel, J.-L. Rigaud, J.-L. Ranck, High
  resolution topographs of the rubrivivax gelatinosus light-harvesting complex
  2., EMBO J. 20 (2001) 3029--3035.

\bibitem{HART2002}
N.~Hartigan, H.~Tharia, F.~Sweeney, A.~Lawless, M.~Papiz, The 7.5-\aa electron
  density and spectroscopic properties of a novel low-light b800 lh2 from
  rhodopseudomonas palustris., Biophys J. 82 (2002) 963--977.

\bibitem{SCHE2004A}
S.~Scheuring, J.~Rigaud, J.~Sturgis, Variable {LH2} stoichiometry and core
  clustering in native membranes of \emph{Rhodospirillum photometricum}., EMBO
  J. 23 (2004) 4127--4133.

\bibitem{JAMI2002}
S.~Jamieson, P.~Wang, P.~Qian, J.~Kirkland, M.~Conroy, C.~Hunter, P.~Bullough,
  Projection structure of the photosynthetic reaction centre-antenna complex of
  \emph{Rhodospirillum rubrum} at 8.5 $\aa$ resolution., EMBO J. 21 (2002)
  3927--3935.

\bibitem{SCHE2003A}
S.~Scheuring, J.~Seguin, S.~Marco, D.~L\'{e}vy, C.~Breyton, B.~Robert,
  J.~Rigaud, {AFM} characterization of tilt and intrinsic flexibility of
  \emph{Rhodobacter sphaeroides} light harvesting complex 2 ({LH2})., J. Mol.
  Biol. 325 (2003) 569--580.

\bibitem{ROSZ2003}
A.~Roszak, T.~Howard, J.~Southall, A.~Gardiner, C.~Law, N.~Isaacs, R.~Cogdell,
  Crystal structure of the {RC-LH1} core complex from \emph{Rhodopseudomonas
  palustris}., Science 302 (2003) 1969--1672.

\bibitem{JUNG99}
C.~Jungas, J.-L. Ranck, J.-L. Rigaud, P.~Joliot, A.~Vermeglio, Supramolecular
  organization of the photosynthetic apparatus of rhodobacter sphaeroides, EMBO
  J. 18 (1999) 534--542.

\bibitem{PARK88}
P.~{Parkes-Loach}, J.~Sprinkle, P.~Loach, Reconstitution of the {B873}
  light-harvesting complex of rhodospirillum rubrum from the separately
  isolated $\alpha$ and $\beta$ polypeptides and bacterichlorophyll $a$,
  Biochemistry 27 (1988) 2718--2727.

\bibitem{TODD98}
J.~Todd, P.~{Parkes-Loach}, J.~Leykam, P.~Loach, In vitro reconstitution of the
  core and peripheral {Light-Harvesting} complexes of \emph{Rhodospirillum
  molischianum} from separately isolated components., Biochemistry 37 (1998)
  17458--17468.

\bibitem{TODD99}
J.~Todd, P.~Recchia, P.~{Parkes-Loach}, J.~Olsen, G.~Fowler, P.~McGlynn,
  C.~Hunter, P.~Loach, Minimal requirements for {in vitro} reconstitution of
  the structural subunit of light-harvesting complexes of photosythesis
  bacteria, Photosynth. Res. 62 (1999) 85--98.

\bibitem{KEHO98}
J.~Kehoe, K.~Meadows, P.~Parkes-Loach, P.~Loach, {Reconstitution of Core
  Light-Harvesting Complexes of Photosynthetic Bacteria Using Chemically
  Synthesized Polypeptides. 2. Determination of Structural Features That
  Stabilize Complex Formation and Their Implications for the Structure of the
  Subunit Complex}., Biochemistry 37 (1998) 3418--3428.

\bibitem{MEAD98}
K.~Meadows, P.~Parkes-Loach, J.~Kehoe, P.~Loach, Reconstitution of core
  light-harvesting complexes of photosynthetic bacteria using chemically
  synthesized polypeptides. 1. minimal requirements for subunit formation.,
  Biochemistry 37 (1998) 3411--3417.

\bibitem{DAVI97}
C.~Davis, P.~Bustamante, J.~Todd, P.~{Parkes-Loach}, P.~McGlynn, J.~Olsen,
  L.~McMaster, P.~Hunter, C.N.and~Loach, Evaluation of structure-function
  relationships in the core light- harvesting complex of photosynthetic
  bacteria by reconstitution with mutant polypeptides, Biochemistry 12 (1997)
  3671--3679.

\bibitem{PARK2004}
P.~{Parkes-Loach}, A.~Majeed, C.~Law, P.~Loach, Interactions stabilizing the
  structure of the core light-harvesting complex ({LH1}) of photosynthetic
  bacteria and its subunit ({B820})., Biochemistry 43 (2004) 7003--7016.

\bibitem{OLSE2003}
J.~Olsen, B.~Robert, C.~Siebert, P.~Bullough, C.~Hunter, Role of the c-terminal
  extrinsic region of the alpha polypeptide of the light-harvesting 2 complex
  of rhodobacter sphaeroides: a domain swap study, Biochemistry 42 (2003)
  15114--15123.

\bibitem{BRAU2002}
P.~Braun, J.~Olsen, B.~Strohmann, C.~Hunter, H.~Scheer, Assembly of
  {Light-Harvesting} bacteriochlorophyll in a model transmembrane helix in its
  natural environment., J. Mol. Biol. 318 (2002) 1085--1095.

\bibitem{koepke96-581}
J.~Koepke, X.~C. Hu, C.~Muenke, K.~Schulten, H.~Michel, The crystal structure
  of the light-harvesting complex ii (b800-850) from rhodospirillum
  molischianum, Structure 4~(5) (1996) 581--597.

\bibitem{prince97-412}
S.~M. Prince, M.~Z. Papiz, A.~A. Freer, G.~McDermott, A.~M.
  Hawthornthwaite-Lawless, R.~J. Cogdell, N.~W. Isaacs, Apoprotein structure in
  the lh2 complex from rhodopseudomonas acidophila strain 10050: modular
  assembly and protein pigment interactions, J. Mol. Biol. 268~(2) (1997)
  412--423.

\bibitem{frenkel2002}
D.~Frenkel, B.~Smit, Understanding Molecular Simulation From Algorithms to
  Applications, Academic Press, California, 2002.

\bibitem{risken96}
H.~Risken, The Fokker-Planck Equation: Methods of Solution and Applications,
  3rd Edition, Springer-Verlag Telos, 1996.

\bibitem{roux95-275}
B.~Roux, The calculation of the potential of mean force using
  computer-simulations, Comput. Phys. Commun. 91~(1-3) (1995) 275--282.

\bibitem{heymann01-1295}
D.~Frenkel, B.~Smit, Extension to the weighted histogram analysis method:
  combining umbrella sampling with free energy calculations, Biophys. J. 81~(3)
  (2001) 1295--1313.

\bibitem{kumar92-1011}
S.~Kumar, D.~Bouzida, R.~H. Swendsen, P.~A. Kollman, J.~M. Rosenberg, The
  weighted histogram analysis method for free-energy calculations on
  biomolecules .1. the method, J. Comput. Chem. 13~(8) (1992) 1011--1021.

\bibitem{isralewitz01-13}
B.~Isralewitz, J.~Baudry, J.~Gullingsrud, D.~Kosztin, K.~Schulten, Steered
  molecular dynamics investigations of protein function, J. Mol. Graphics
  Modell. 19~(1) (2001) 13--25.

\bibitem{grubmuller96-997}
H.~Grubmuller, B.~Heymann, P.~Tavan, Ligand binding: Molecular mechanics
  calculation of the streptavidin biotin rupture force, Science 271~(5251)
  (1996) 997--999.

\bibitem{jarzynski97-2690}
C.~Jarzynski, Nonequilibrium equality for free energy differences, Phys. Rev.
  Lett. 78~(14) (1997) 2690--2693.

\bibitem{park04-5946}
S.~Park, K.~Schulten, Calculating potentials of mean force from steered
  molecular dynamics simulations, J. Chem. Phys. 120~(13) (2004) 5946--5961.

\bibitem{hummer01-3658}
G.~Hummer, A.~Szabo, Free energy reconstruction from nonequilibrium
  single-molecule pulling experiments, Proc. Natl. Acad. Sci. U. S. A. 98~(7)
  (2001) 3658--3661.

\bibitem{gore03-12564}
J.~Gore, F.~Ritort, C.~Bustamante, Bias and error in estimates of equilibrium
  free-energy differences from nonequilibrium measurements, Proc. Natl. Acad.
  Sci. U. S. A. 100~(22) (2003) 12564--12569.

\bibitem{schwieters03-66}
C.~D. Schwieters, J.~J. Kuszewski, T.~N, G.~M. Clore, The xplor-nih nmr
  molecular structure determination package, J. Magn. Res. 160 (2003) 66--74.

\bibitem{humphrey96-33}
W.~Humphrey, A.~Dalke, K.~Schulten, {VMD}: Visual molecular dynamics, J. Mol.
  Graph. 14~(1) (1996) 33.

\bibitem{damjanovic02-031919}
A.~Damjanovic, I.~Kosztin, U.~Kleinekathöfer, K.~Schulten, Excitons in a
  photosynthetic light-harvesting system: a combined molecular dynamics,
  quantum chemistry, and polaron model study, Phys. Rev. E 65 (2002) 031919.

\bibitem{kale99-283}
L.~Kale, R.~Skeel, M.~Bhandarkar, R.~Brunner, A.~Gursoy, N.~Krawetz,
  J.~Phillips, A.~Shinozaki, K.~Varadarajan, K.~Schulten, Namd2: greater
  scalability for parallel molecular dynamics, J. Comput. Phys. 151~(1) (1999)
  283--312.

\bibitem{mackerell98-3586}
A.~D. MacKerell~Jr, D.~Bashford, M.~Bellott, R.~L. Dunbrack~Jr, J.~Evanseck,
  M.~J. Field, S.~Fischer, J.~Gao, H.~Guo, S.~Ha, D.~Joseph, L.~Kuchnir,
  K.~Kuczera, F.~T.~K. Lau, C.~Mattos, S.~Michnick, T.~Ngo, D.~T. Nguyen,
  B.~Prodhom, I.~W.~E. Reiher, B.~Roux, M.~Schlenkrich, J.~Smith, R.~Stote,
  J.~Straub, M.~Watanabe, J.~Wiorkiewicz-Kuczera, D.~Yin, M.~Karplus,
  All-hydrogen empirical potential for molecular modeling and dynamics studies
  of proteins using the charmm22 force field, J. Phys. Chem. B 102 (1998)
  3586--3616.

\bibitem{PAND2001}
A.~Pandit, R.~Vischers, I.~{van Stokkum}, R.~Kraayenhof, R.~{van Grondelle},
  Oligomerization of {Light-Harvesting} {I} antenna peptides of
  \emph{Rhodospirillum rubrum}, Biochemistry 40 (2001) 12913--12924.

\bibitem{VEGH2002}
A.~V\'{e}gh, B.~Robert, Spectroscopic characterisation of a tetrameric subunit
  form of the core antenna protein from \emph{Rhodospirillum rubrum}., FEBS
  Lett. 528 (2002) 222--226.

\bibitem{SCHU2004}
A.~Schubert, A.~Stenstam, W.~J.~D. Beenken, J.~L. Herek, R.~Cogdell,
  T.~Pullerits, V.~Sundstrom, In vitro self-assembly of the light harvesting
  pigment-protein lh2 revealed by ultrafast spectroscopy and electron
  microscopy, Biophys. J. 86~(4) (2004) 2363--2373.

\end{thebibliography}

\end{document}